\documentstyle[prc,eqsecnum,aps,psfig]{revtex}

\begin{document}
\draft

\twocolumn[\hsize\textwidth\columnwidth\hsize\csname
@twocolumnfalse\endcsname

\title{Continued fraction representation of the Coulomb
Green's operator \\
and unified  description of bound, resonant and scattering states} 

\author{B. K\'onya$^{1,2}$, G. L\'evai$^1$, and Z. Papp$^{1,2}$}
\address{$^1$ Institute of Nuclear Research of the Hungarian 
Academy of Sciences, \\
P.O. Box 51, H--4001 Debrecen, Hungary \\
$^2$ Institute for Theoretical Physics, University
of Graz,\\ Universit\"atsplatz 5, A-8010 Graz, Austria}
\date{\today}
\maketitle
\begin{abstract}
If a quantum mechanical Hamiltonian 
has an infinite symmetric tridiagonal (Jacobi) matrix form
in some discrete Hilbert-space basis representation, then  
its Green's operator can be constructed in terms of a continued fraction.
As an illustrative example we discuss the Coulomb Green's operator in
Coulomb--Sturmian basis representation.
Based on this representation, a quantum mechanical approximation
method for solving Lippmann--Schwinger integral equations
can be established, which is equally applicable for bound-, resonant-
and scattering-state problems with free and Coulombic asymptotics as well.
The performance of this technique is illustrated with a detailed 
investigation of a nuclear potential describing 
the interaction of two $\alpha$ particles. 
\end{abstract}
\vspace{0.5cm}


]

\narrowtext

\section{Introduction}

Green's operators play a central role in theoretical physics, especially
in quantum mechanics, as they appear in fundamental equations governing
the dynamics of physical systems. The formalisms based on Green's
operators represent alternative, but essentially equivalent way of
describing the same systems as methods based on differential equations.
However, since Green's operators are related to the integral equation 
formalism, their use  is more advantageous in many cases than that of 
traditional differential approaches: boundary conditions, for example, 
are automatically incorporated in the formalism. Although the concept 
of Green's operators frequently appears in basic textbooks on the level 
of fundamental equations, in practical calculations their use is usually 
avoided, and various approximations to the Schr\"odinger equation are 
preferred instead. The reason certainly is that the evaluation of Green's 
operators is much more complicated than the direct treatment of the 
Hamiltonian using standard tools of theoretical and mathematical physics.

The discrete Hilbert-space basis representation of the Green's operator 
is very advantageous since it makes the solution of the 
Lippmann--Schwinger-type integral equations possible on 
an easy-to-apply, yet very general way. If the integral equation possesses
good mathematical properties, then the potential term can be well approximated 
on a finite subset of the Hilbert-space basis and the integral equation can 
be solved without any further approximation.
This concept has already been elaborated in several different forms.
In Refs.\ \cite{revai,hopse} harmonic oscillator functions were used, which 
allowed the representation of the free Green's operator, and thus the 
corresponding approximation method can handle problems with
free asymptotics.  In Refs.\  
\cite{papp1,papp2,papp3,cpc} Coulomb--Sturmian basis was applied, which 
allowed the inclusion of the long-range Coulomb potential into the 
Green's operator, and thus led to the exact treatment of the Coulomb 
asymptotics. This latter approach has also been extended to solving the 
three-body Coulomb problem in the Faddeev approach. So far good results 
have been reached for bound-state \cite{pzwp} and below-breakup 
scattering-state problems \cite{pzsc}. These results have showed the 
efficiency of the discrete Hilbert-space expansion method in solving 
fundamental integral equations. We note that
in all the previous approaches \cite{papp1,papp2,papp3,cpc,pzwp,pzsc} 
the Coulomb Green's matrix has 
been determined in 
a rather complicated way with the aid of special functions.

In Ref.\ \cite{jmp} we have presented a rather general and easy-to-apply
method for calculating the discrete Hilbert-space basis 
representation of the Green's 
operators of those Hamiltonians, which have infinite symmetric 
tridiagonal (i.e.\ Jacobi) matrix forms. The procedure necessitates the
evaluation of the Hamiltonian matrix on 
this basis, a rather common task in quantum mechanics. 
Then the elements of the Jacobi matrix are 
used as input in the calculation of the Green's matrix in terms of 
a continued fraction.
This way of calculating Green's matrices 
simplifies the calculations considerably and the representation
via continued fraction provides a readily computable way.
The combination of this new way of calculating the Coulomb Green's matrix
with the technique of solving integral equations in discrete
Hilbert--space-basis representation results a quantum mechanical
approximation method which is rather general in the sense that it is 
equally applicable to solving bound-, resonant- and scattering-state 
problems with practically any potential of physical relevance. 
And all this is provided at a very little cost: in practice only
matrix elements of the Hamiltonian are required. 

The structure of this paper is the following. In section \ref{ketto}
we consider the  Coulomb Green's 
operator in Coulomb--Sturmian representation. We show that 
this Coulomb Green's matrix can be given in terms of a continued
fraction. In section \ref{harom} 
the solution method of the Lippmann--Schwinger equation in 
Coulomb-Sturmian space representation is explained, 
and the method is applied to describe 
bound, resonance and scattering solutions of a model nuclear potential. 
Finally conclusions are drawn in section \ref{conc}.

\section{Continued fraction representation of the Coulomb Green's operator}
\label{ketto}

Let us consider the radial Coulomb Hamiltonian 
\begin{equation}
 H^{\rm C}_l=-\frac{\hbar^2}{2m}\left(\frac{\mbox{d}^2 }{\mbox{d} r^2}
+ \frac{l(l+1)}{ r^2}\right) + \frac{Z{\rm e}^2}{ r}\ ,
\label{coulham}
\end{equation}
where $m$, $l$, ${\rm e}$ and $Z$ stand for the mass, angular momentum,
electron charge and charge number, respectively.
The Coulomb--Sturmian (CS) functions, the Sturm--Liouville
solutions of the Hamiltonian (\ref{coulham}) \cite{rotenberg}, appear as
\begin{equation}
\langle r\vert n  \rangle 
= \sqrt{ \frac{n!}{(n+2l+1)!} } \ 
\exp(-b r) (2b r)^{l+1} L_n^{(2l+1)}(2b r)\ , 
\label{csf}
\end{equation}
where $b$ is a scale parameter, $n$ is the radial quantum number and 
$L_n^{(\alpha)}$ denotes the generalized Laguerre polynomials \cite{as}. 
With the biorthonormal partner defined by 
$\langle r\vert \widetilde{ n } \rangle \equiv
\langle r\vert n  \rangle/r$ these functions form a discrete basis.
We denote the Coulomb Green's operator
as $G^{\rm C}_l (z)=(z-H^{\rm C}_l)^{-1}$ and we consider here its CS
matrix elements $\underline{G}^{\rm C}_{n n'}=
\langle \widetilde{ n  } | G^{\rm C}_l |
\widetilde{ n'   } \rangle$.

The starting point in this procedure is the observation the
the matrix 
$\underline{J}^{\rm C}_{n n^{\prime}}=
\langle n |(z- H^{\rm C}_l)|n^{\prime } \rangle$ 
possesses an infinite symmetric tridiagonal i.e.\ Jacobi structure,
\begin{equation}
\underline{J}^{\rm C}_{nn}=2(n+l+1) (k^2-b^2 )
\frac{\hbar^2}{4mb}- Z{\rm e}^2 
\label{jii}
\end{equation}
and
\begin{equation}
\underline{J}^{\rm C}_{nn-1}=-[n(n+2l+1)]^{1/2} (k^2+b^2 )
\frac{\hbar^2}{4mb} \ , 
\label{jiip1}
\end{equation}
where $k=(2m z/\hbar^2)^{1/2}$ is the wave number.
The main result of Ref.\ \cite{jmp} is that for Jacobi matrix systems 
the $N$'th leading submatrix 
$\underline{G}^{{\rm C} (N)}_{ij}$ of the infinite Green's matrix
can be determined by the elements of the Jacobi matrix
\begin{equation}
\label{invn}
\underline{G}^{{\rm C} (N)}_{ij}=[\underline{J}^{\rm C}_{ij}+ 
\delta _{jN}\, \, 
\delta _{iN}\, \, \underline{J}^{\rm C}_{NN+1}\, 
\, C ]^{-1}\ ,
\end{equation}
where $C$ is a  continued fraction
\begin{equation}
C=-\frac{u_N}{d_N+
\frac{\displaystyle u_{N+1}}{\displaystyle d_{N+1}+ 
\frac{\displaystyle u_{N+2}}{ \displaystyle d_{N+2} + \cdots }}} \ ,
\label{frakk}
\end{equation}
with coefficients 
\begin{equation}
u_i=- {\underline{J}^{\rm C}_{i,i-1}}/{\underline{J}^{\rm C}_{i,i+1}}, 
\quad d_i=- {\underline{J}^{\rm C}_{i,i}}/{\underline{J}^{\rm C}_{i,i+1}} 
\ .
\label{egyutth}
\end{equation}

In Ref.\ \cite{jmp} we have shown that the
continued fraction $C$, as it stands, 
is convergent only for negative energies, 
but it can be continued analytically to the whole complex energy plane.
Since the $u_i$ and $d_i$ coefficients possess the limit properties
\begin{equation}
u \equiv \lim_{i\rightarrow \infty }u_i=-1 
\end{equation}
and
\begin{equation}
d\equiv \lim_{i\rightarrow \infty} d_i= 2(k^2 -   b^2)/ 
 ( k^2 +   b^2)\ , 
\label{ud}
\end{equation}  
the continued fraction  appears as
\begin{equation}
C=-\frac{u_N}{d_N+
\frac{\displaystyle u_{N+1}}{\displaystyle d_{N+1}+ \cdots +
\frac{\displaystyle u}{ \displaystyle d +  
\frac{\displaystyle u}{ \displaystyle d +    \cdots 
 }}}}\ .
\label{cflim}
\end{equation}
The tail  $w$ of $C$ satisfies the implicit
relation
\begin{equation}
w=\frac{u}{d+w}\ , 
\label{tail2}
\end{equation}
which is solved by
\begin{equation}
\label{wpm}
w_{\pm}=(b \pm  {\mathrm{i}}k )^2/(b^2+k^2)  \ .
\end{equation}
Replacing the tail of the continued fraction by its explicit analytical 
form $w_{\pm}$, we can speed up the convergence and, which is more important, 
turn a non-convergent continued fraction into a
convergent one \cite{lorentzen}. 
In fact, by using $w_{\pm}$  instead of the non-converging tail
we perform an analytic continuation. 
In Ref.\ \cite{jmp} we have 
shown that $w_+$ provides an analytic continuation of the Green's matrix 
to the physical, while $w_{-}$ to the unphysical Riemann-sheet. This way 
Eq.\ (\ref{frakk}) together with (\ref{invn}) provide the CS basis
representation of the Coulomb Green's operator on the whole complex
energy plane.  We note here that with the choice 
of $Z=0$ the Coulomb Hamiltonian (\ref{coulham}) reduces to the kinetic 
energy operator and our formulas provide the CS basis representation of the 
Green's operator of the free particle as well.

\begin{figure}
\psfig{file=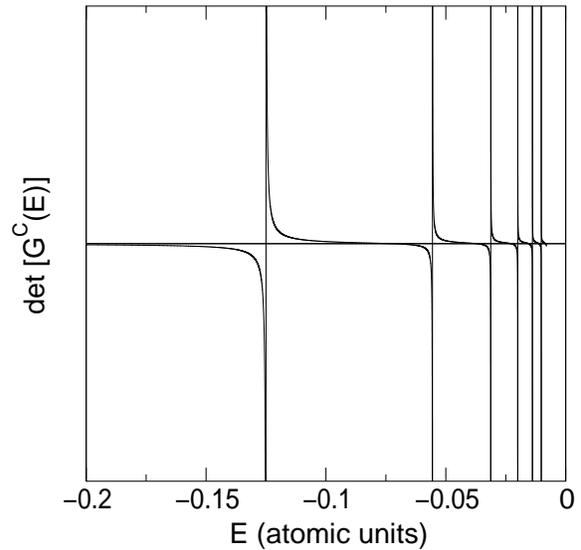,width=7.5cm,angle=-90}
\caption{The determinant of a $3 \times 3 $  Coulomb Green's matrix
${\rm det} [ \underline{G}^C(E)]$ as the function of
the energy $E$ for $l=1$.
The bound states of the Coulomb problem are
located at energies where the vertical lines cross the 
horizontal axis. (These lines are shown only for demonstrative
purposes: they do not correspond to functional values of
${\rm det} [ \underline{G}^C(E)]$.) Atomic units of $m={\rm e}=\hbar=1$ 
and $Z=-1$ were used.
For the sake of clarity only the first 6 energy levels are shown.
These are located at $E_n=-1/[2(n+l+1)]^2$ with $n\le 5$. 
}
\label{coulfig}
\end{figure}

We can immediately test  $\underline{G}^C$ by calculating 
eigenvalues belonging to attractive Coulomb interactions. 
Fig.\ \ref{coulfig} shows
$|{\underline{G}^C}(E)|$.  The poles coincide with the exact
Coulomb energy levels 
up to machine accuracy. We stress that, from the point of view of 
determining the energy eigenvalues, the rank $N$ of the matrix 
and the specific choice of the CS basis parameter are irrelevant.
Any representation of the Coulomb Green's operator exhibits all the properties
of the system and our Green's matrix contains all the infinitely many
eigenvalues. This is especially interesting if we compare with the usual
procedure of calculating eigenvalues of a finite Hamiltonian matrix which could
only provide an upper limit for the N lowest eigenvalues. Our procedure 
does not truncate the Coulomb Hamiltonian, because
all the higher $\underline{J}_{nn'}$  matrix elements are implicitly contained 
in the continued fraction. We note that $\underline{G}^C$ has already
been calculated before \cite{papp1,papp2,papp3,cpc} by using $\mbox{}_2F_1$
hypergeometric functions \cite{as}.  In Ref.\ \cite{jmp} we have shown that 
the two formalism lead to numerically identical results for all energies and
the continued fraction representation possesses all the analytic properties 
of $G^C$ also in practice.

\section{Solution of the Lippmann-Schwinger 
integral equation in CS representation}
\label{harom}

In this section, after Refs.\ \cite{papp1,papp2,papp3,cpc},
we recapitulate the solution of the Lippmann--Schwinger 
equation in CS 
representation. We suppose that the Hamiltonian $H_l$ is split into
two terms  
\begin{equation}
H_l=H_l^{\rm C} + V_l \ ,
\end{equation}
where $V_l$ is the asymptotically irrelevant short-range potential. 
The Green's operator of  $H_l$  is defined by 
$G_l(z)=(z-H_l)^{-1}$,  and 
it is connected to $G_l^{\rm C}(z)$ via the resolvent relation 
\begin{equation}
G_l(z)=G_l^{\rm C}(z)+G_l^{\rm C} (z)V_l G_l(z)\ .
\label{resolvent}
\end{equation}
The wave function $|\Psi^{(\pm)}_l \rangle$ describing a 
scattering process satisfies the inhomogeneous Lippmann--Schwinger 
equation \cite{newton}
\begin{equation}
|\Psi^{(\pm)}_l  \rangle =|\Phi^{(\pm)}_l \rangle 
+G_l^{\rm C}(E \pm {\rm i}0) V_l |\Psi^{(\pm)}_l \rangle \ ,  
\label{LS}
\end{equation}
where $|\Phi^{(\pm)}_l \rangle $ is the solution of the Hamiltonian 
$H^{\rm C}_l$ possessing scattering asymptotics.  The bound- and 
resonant-state  wave functions 
satisfy the homogeneous Lippmann--Schwinger  equation
\begin{equation}
|\Psi_{l }  \rangle = G_l^{\rm C}(E ) V_l |\Psi_{l }\rangle  
\label{LS1}
\end{equation}
at real negative and complex $E$ energies, respectively.

We are going to solve these equations in a unified way by approximating
only the potential term $V_l$. For this purpose we write the unit operator
in the form 
\begin{equation}
\label{biort3}
{\bf 1_l}=\lim_{N\to\infty} {\bf {1}}_{l N}\ ,
\end{equation}
where 
\begin{equation}
{\bf {1}}_{l N}= \sum _{n=0}^{N }|\tilde{n}\rangle \sigma_n^N
\langle n|=\sum _{n=0}^{N }|n\rangle \sigma_n^N \langle \tilde{n}|\ .
\label{1N}
\end{equation}
The $\sigma$ factors have the properties $\lim_{n\to\infty} \sigma_n^N =0$
and $\lim_{N\to\infty} \sigma_n^N =1$, and make the limit in (\ref{biort3})
smoother. They were introduced originally for improving the convergence 
properties of truncated trigonometric series \cite{lanczos}, 
but they turned out to be also very efficient  in solving integral equations 
in discrete Hilbert space basis representation \cite{borbely}. 
The choice of $\sigma_n^N$ 
\begin{equation}
\sigma_n^N = \frac{1-\exp\{-[\alpha(n-N-1)/(N+1)]^2\}}{1-\exp(-\alpha^2)}
\label{sigma}
\end{equation}
with $\alpha\sim 5$ has proved to be appropriate in practical calculations. 

Let us introduce an approximation of the potential operator 
\begin{equation}
V_l = {\bf 1_l} V_l {\bf 1_l} \approx {\bf {1}}_{l N} V_l 
{\bf {1}}_{l N} = V_l^N = 
\sum_{n,n' =0}^N
|\widetilde{n}\rangle  \;
\underline{V}_{n n^\prime} \;\mbox{} \langle \widetilde{n^{\prime }
}| \ ,
\label{sepfe2b}
\end{equation}
where the matrix elements
\begin{equation}
 \underline{V}_{n n^\prime} =
\sigma^N_{n} \langle n|
V_l| n^{\prime } \rangle \sigma^N_{n^{\prime}}\ ,  
\label{v2b}
\end{equation}
in general, have to be calculated numerically.
This approximation is called separable expansion, because the operator
$V_l^N$, { e.g.}\ in coordinate representation, appears in the form 
\begin{equation}
\langle r | V^N| r^{\prime } \rangle = \sum_{n, n' =0}^N
\langle r |\widetilde{n }\rangle  \;
 \underline{V}_{n n'} \;\mbox{} \langle \widetilde{n^{\prime }
}| r^{\prime} \rangle\ , 
\end{equation}
i.e.\  the dependence on $r$ and $r^{\prime}$ appears in a separated 
functional form.

With this  separable potential Eqs.\ (\ref{LS}) and  (\ref{LS1}) reduce to 
\begin{equation}
|\Psi^{(\pm)}_l \rangle =|\Phi^{(\pm)}_l \rangle + \sum_{n,n^\prime =0}^N
G^{\rm C}_l (E\pm {\rm i}0)  |\widetilde{n}\rangle  \;
\underline{V}_{n n'} \;\mbox{} \langle \widetilde{n^{\prime }}
|\Psi^{(\pm)}_l \rangle\ ,  
\label{LSapp1}
\end{equation}
and 
\begin{equation}
|\Psi_{l } \rangle = \sum_{n,n^\prime =0}^N
G_l^{\rm C}(E ) |\widetilde{n}\rangle  \;
 \underline{V}_{n n'}\;\mbox{} \langle \widetilde{n^{\prime }}
|\Psi_{l} \rangle\ ,  
\label{LSapp2}
\end{equation}
respectively. To derive equations for the 
coefficients $\underline{\Psi}_l^{(\pm)} =
\langle \widetilde{n^{\prime }}
|\Psi_l^{(\pm)} \rangle$ and $\underline{\Psi }_l 
=\langle \widetilde{n^{\prime }}
|\Psi_{l} \rangle$, we have to
act with states $\langle \widetilde{n''}|$
from the left. Then the following inhomogeneous and homogeneous 
algebraic equations are obtained, for scattering and bound-state 
problems, respectively: 
\begin{equation}
\lbrack (\underline{G}^{\rm C}_l (E \pm {\rm i}0))^{-1}-
\underline{V}_l]\underline{\Psi}_l^{(\pm)}=
\underline{\Phi}_l^{(\pm)},  \label{eq18a}
\end{equation}
and
\begin{equation}
\lbrack (\underline{G}_l^{\rm C} (E))^{-1}-
\underline{V}_l]\underline{\Psi }_l= 0\ ,
  \label{eq18b}
\end{equation}
where the overlap
${\Phi}_{n l}^{(\pm)}=\langle  \widetilde{n}|\Phi_l^{(\pm)} \rangle $
can also be calculated analytically \cite{papp2}.
 The homogeneous equation (\ref{eq18b}) is solvable if and only if 
\begin{equation}
\label{det}
\det [(\underline{G}_l^{\rm C}(E))^{-1}-\underline{V}_l]=0 
\label{determinans}
\end{equation}
holds, which is an implicit nonlinear equation for the bound- and
resonant-state energies.  
As far as the scattering states concerned the solution of (\ref{eq18a}) 
provides the overlap $\langle \widetilde{n}|\Psi_l \rangle$. 
From this quantity
any scattering information can be inferred, for example the
scattering amplitude corresponding to potential $V_l$ is given by 
\cite{newton}
\begin{equation}
\label{scamp}
 A^{V} = \langle 
\Phi^{(-)}_l | V_l | \Psi^{(+)}_l  \rangle = \underline{\Phi}_l^{(-)}
\underline{V}_l
 \;\underline{\Psi}_l^{(+)}.
\end{equation}
Note that also the Green's matrix of the total Hamiltonian,
which is equivalent to the complete solution of the physical system,
can be constructed as the solution of Eq.\ (\ref{resolvent})
\begin{equation}
\underline{G}_l(z)=\lbrack (\underline{G}_l^{\rm C} (z))^{-1}-
\underline{V}_l ]^{-1} \ .
\label{totgrm}
\end{equation}

Finally, it should also be emphasized, that in this approach only the
potential operator is approximated, but the asymptotically important 
$H^{\rm C}$
term remains intact. The solutions are defined on the whole Hilbert 
space, not only on a finite subspace. The wave functions are not
linear combinations of basis functions, but rather, as Eqs.\ (\ref{LSapp1})
and (\ref{LSapp2}) indicate, linear combinations of the states
$G_l^{\rm C}(E)  |\widetilde{n}\rangle$, which have been shown to possess 
correct Coulomb asymptotics \cite{papp3}.

\subsection{Bound, resonant and scattering states
in a model nuclear potential}

\label{harom-harom}

In this subsection we apply the above techniques to 
calculate bound-, resonant- and scattering-state
solutions of potential problems. The 
particular example we consider is a potential representing  the 
interaction of two $\alpha$ particles. 
This example is thoroughly discussed in the pedagogical work 
\cite{schmid} in the context of a conventional approach based on 
the numerical solution of the Schr\"odinger equation. 
The interaction of  two $\alpha$ particles can be approximated 
by the potential  
\begin{equation}
\label{be8pot}
{V}_{\alpha-\alpha}(r)=
 -A \exp(-\beta r^2) + \frac{Z^2{\rm e}^2}{r}\ {\rm erf}(\gamma r) , 
\end{equation}
where ${\rm erf} (z)$ is the error function \cite{as}. 
This potential is a composition of a bell-shaped deep, attractive 
nuclear potential, and a repulsive electrostatic field between two 
extended  charged objects. 
The units used in the Hamiltonian of this system 
are suited to nuclear physical applications,  i.e.\ 
the energy and length scale are measured in MeV
and fm, respectively. In these units 
$\hbar/(2m)=10.375$ MeV fm$^2$ (with $m$ being the reduced mass of 
two $\alpha$ particles) and ${\rm e}^2= 1.44$ MeV fm. The other 
parameters are $A=$122.694 MeV, $\beta=0.22$ fm$^{-2}$, $\gamma=0.75$ 
fm$^{-1}$  and $Z=2$ (the charge number of the $\alpha$ 
particles). 

Since the $\alpha-\alpha$ potential possesses a Coulomb tail,
for the   asymptotically relevant operator we should take 
the Coulomb Hamiltonian $H^{\rm C}$ 
of Eq.\ (\ref{coulham}) and the short-range potential is 
then defined by 
\begin{equation}
 V(r)=V_{\alpha-\alpha}(r)-\frac{Z^2{\rm e}^2}{r} = 
 -A \exp(-\beta r^2) - \frac{Z^2{\rm e}^2}{r}\ {\rm erfc}(\gamma r) ,
\end{equation} 
with ${\rm erfc} (z)=1- {\rm erf} (z)$. 
We approximate the short-range
potential $V(r)$ according to Eq.\ (\ref{sepfe2b}) on the CS basis. 
From this point on, however, the properties
of the physical system will be buried into the numerical values of
the matrix elements. Thus the method is  applicable to all types of
potentials, as long as we can calculate their matrix elements somehow. 
Besides ordinary potentials this equally applies to complex, 
momentum-dependent, non-local, etc.\  potentials relevant to 
practical problems of atomic, nuclear and particle physics. 
Furthermore, the present formalism 
is equally suited to problems including attractive or repulsive 
long-range Coulomb-like  and short-range potentials. 

It should also be emphasized again that in this approach only
the potential operator $V$ is approximated and the resulting wave 
function possesses the correct asymptotics. So, if the parameter $b$ 
is chosen such that $V^N$ in (\ref{sepfe2b}) 
approaches $V$ uniformly, the convergence
of the bound-, resonant- and scattering-state quantities with respect to
increasing $N$ will also be uniform, thus $V^N$  provides more or less 
uniformly good approximation to $V$ on the whole spectrum of 
physical interest.

\subsubsection{Bound states}

First we  consider only the nuclear part of potential (\ref{be8pot}) 
and switch off the Coulomb interaction by setting $Z=0$.
According to Ref.\ \cite{schmid}, this potential supports altogether 
four bound states: three with $l=0$ and one with $l=2$. 
However, it is 
known that the first two $l=0$ and the single $l=2$ state are unphysical, 
because they are forbidden due to the Pauli principle. 
This fact is not 
taken into account in this simple potential model. 
Although from the physical point of view 
these Pauli-forbidden states have to be dismissed as unphysical, 
they are legitimate solutions of our simple model
potential. The proper inclusion of the Pauli principle in the model
would turn the potential into a non-local one. This problem has been 
considered within the present method in Ref.\ \cite{papp2}.

As illustrative examples we present the results of our calculations 
for the three $l=0$ states. We determined the energies of these states 
from Eq.\ (\ref{det}), using the CS parameter $b=4$ fm$^{-1}$. 
Table \ref{tab1} shows 
the convergence of the method with respect to $N$, the number of basis 
states used in the expansion. It can be seen that the method is very accurate,
convergence up to 12 digits can easily been reached. 
We note that according to Ref.\ \cite{schmid}, the energy of the two 
lowest (i.e.\ the unphysical) $l=0$ states is $E=-76.903\ 6145$ and 
$-29.000\ 48$ MeV in the uncharged case, which is in reasonable agreement 
with our results.

\subsubsection{Resonance states}

Switching on the repulsive Coulomb interaction ($Z=2$) the bound states 
are shifted to higher energies. The most spectacular effect is that 
the third $l=0$ state, which is located at $E=-1.608\ 740\ 8214$ MeV in 
the uncharged case, moves to positive energies and becomes a resonant state. 
This is in agreement with 
the observations: the $\alpha-\alpha$ system (i.e.\ the $^8$Be nucleus) 
does not have a stable ground state, rather it decays  with a half life 
of $7\times 10^{-17}$ sec.

\begin{figure}
\psfig{file=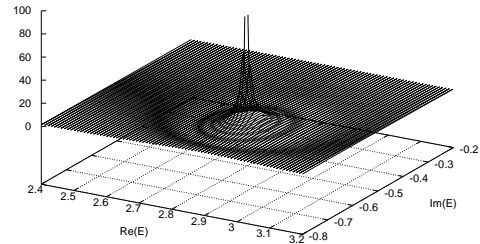,width=7.5cm,angle=-90}
\caption{The modulus of the determinant of the $ \underline{G}(E)$ Green's
matrix for 
the $\alpha-\alpha$ potential on the complex energy plane for $l=2$.
The pole at $E=2.889\ 25-{\rm i\ } 0.620\ 61$ MeV  corresponds 
to a resonance.}
\label{be8res}
\end{figure}

In our calculations we determined the energy corresponding to this 
resonance and to other ones as well by the same techniques we used before 
to find  bound states. In fact, we used the same 
computer code and the same CS parameter ($b=4$ fm$^{-1}$) as we used in the 
analysis of bound states. The method, again, requires locating the poles 
of the Green's matrix, but not on the real energy 
axis, rather on the complex energy plane. 
In Table \ref{tab2} we demonstrate the convergence of our method 
with respect to $N$ for the lowest $l=0$ and $l=2$ resonance states. 
In Fig.\ \ref{be8res} we plotted the modulus of the determinant of the 
Green's matrix (as in Eq.\ (\ref{totgrm}))  
$|\underline{G}(E)|$  over the complex energy plane for $l=2$. 
The resonance is located at the pole of this function. 
Finally, we mention that there is a resonance state for $l=4$ at 
$E=11.791\ 038-{\rm i\ } 1.788\ 957$ MeV.  

Although it is not our aim here to reproduce experimental data 
with this simple potential model, we note that the corresponding 
experimental values \cite{ajzenberg}
are $E_{{\rm r}, l=0}=$0.09189  MeV, 
$E_{{\rm r}, l=2}=$3.132 $\pm$ $0.030$ MeV, $E_{{\rm r}, l=4}=$11.5 $\pm$ 
0.3 MeV, 
and $\Gamma_{l=0}/2 = (3.4 \pm 0.9)\times 10^{-6}$ MeV, 
$\Gamma_{l=2}/2$=0.750 $\pm$ 0.010 MeV, 
$\Gamma_{l=4}/2\simeq$1.75  MeV.

\subsubsection{Scattering states}

In order to demonstrate the performance of our approach for scattering 
states we calculated scattering phase shifts $\delta_l(E)$ for 
${V}_{\alpha-\alpha}(r)$ in (\ref{be8pot}). As described previously in this 
section, phase shifts can be extracted from the scattering amplitude 
given in Eq.\ (\ref{scamp}). Specifying this formula for the Coulomb-like 
case and for a given partial wave $l$ we have 
\begin{equation}
a_l=\frac{1}{k}\exp({\rm i}(2\eta_l+\delta_l))\sin \delta_l\ ,
\label{coulphsh}
\end{equation}
where $a_l$ is the Coulomb-modified scattering amplitude corresponding
to the short-range potential, 
$\eta_l={\rm arg} \Gamma(l+i\gamma+1))$ is the phase shift of the Coulomb 
scattering with $\gamma=Z^2{\rm e}^2m/\hbar^2k$ 
and $\delta_l$ is the phase shift due to the short-range potential. 

The convergence of the method with respect to $N$ is demonstrated in Table 
\ref{tab3}, where $\delta_0(E)$ is displayed at three different energy 
values $E$. As in our calculations for the bound and the resonance states, 
we used $b=4$ fm$^{-1}$here too.  

\begin{figure}
\psfig{file=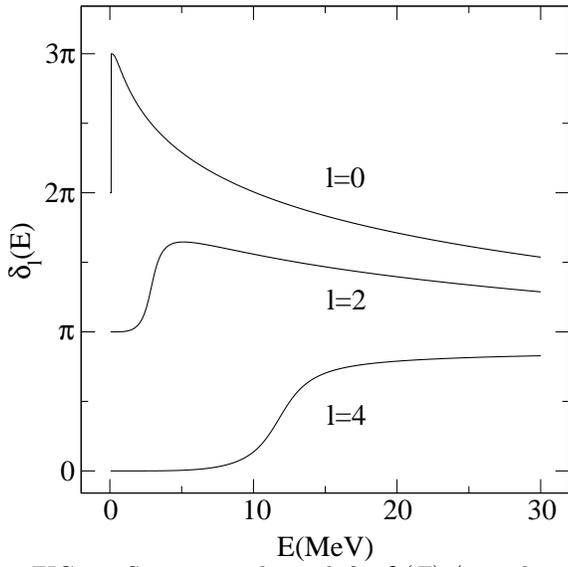,width=7.5cm,angle=-90}
\caption{Scattering phase shifts $\delta_l(E)$ (in radians) 
in the $\alpha-\alpha$ 
potential for $l=0$, 2 and 4. The resonances in these partial waves appear 
as sharp rises in the corresponding phase shifts. In these calculations 
a basis with $N=35$ was used.}
\label{phsh}
\end{figure}

In Fig.\ \ref{phsh} we plotted the scattering phase shifts $\delta_l(E)$ for 
$l=0$, 2 and 4 up to $E=30$ MeV. 
In all three plots in Fig.\ \ref{phsh} the location of the corresponding 
resonance is clearly visible as a sharp rise of the phase around the 
resonance energy $E_{\rm r}$. This rise is expected to be more sudden 
for sharp resonances, and this is, in fact, the case 
here too. 
The phase changes with an abrupt jump of $\pi$ for the sharp $l=0$ resonance, 
while it is slower for the broader $l=2$ and $l=4$ resonances. 
We also note that the phase shifts plotted in Fig.\ \ref{phsh} are also 
in accordance with the Levinson theorem, which states 
that $\delta_l(0)=m\pi$, where $m$ is the number of bound states in the 
particular angular momentum channel. Indeed, as we have discussed earlier, 
there are two bound states for $l=0$, one for $l=2$ and none for $l=4$. 

\begin{figure}
\psfig{file=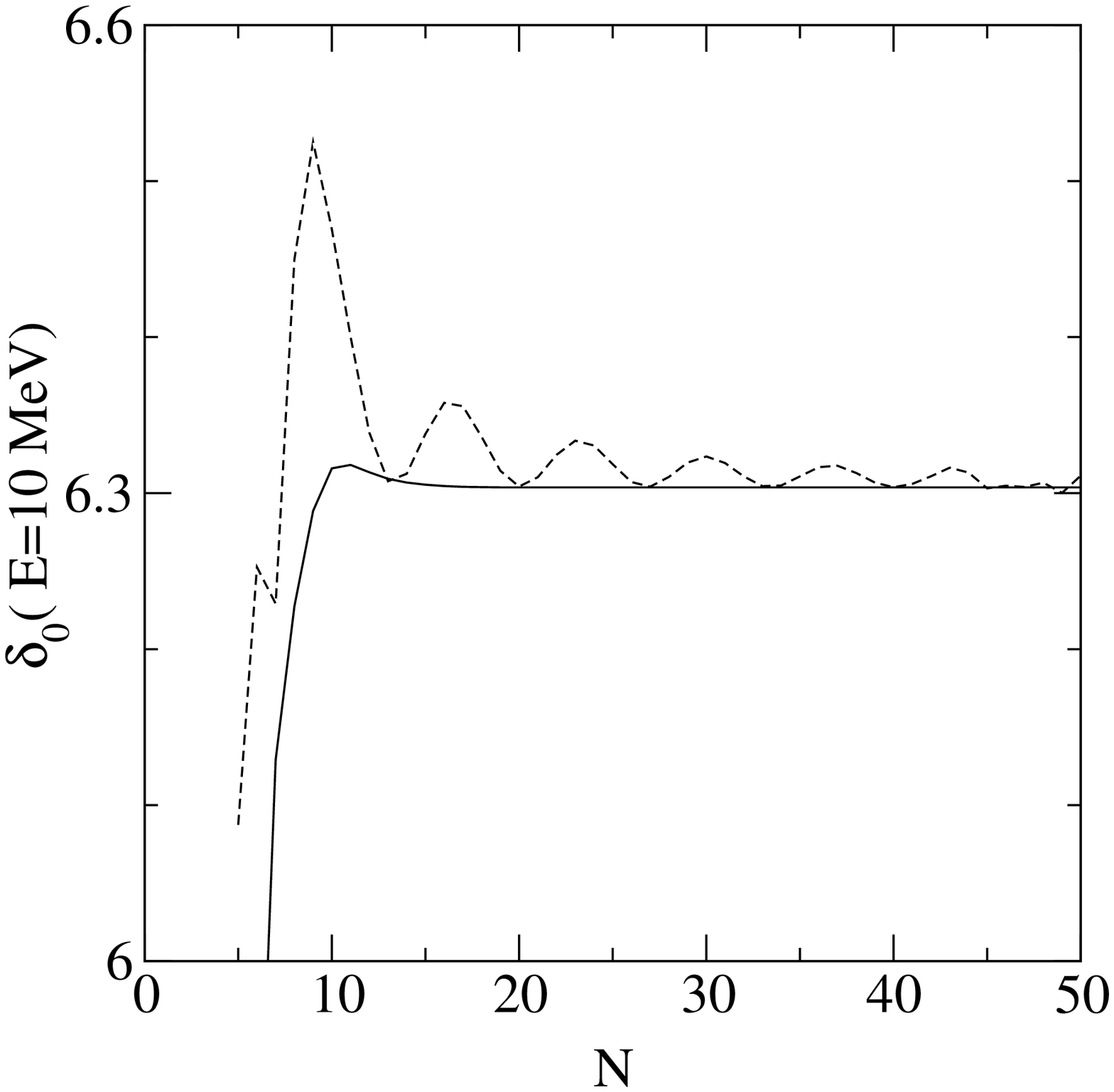,width=7.5cm,angle=-90}
\caption{Convergence of the scattering phase shift $\delta_0(E)$ 
(in radians) at $E=10$ 
MeV calculated with (solid line) and without (dashed line) smoothing factors. }
\label{simicska}
\end{figure}

As an illustration of the importance of the smoothing factors 
we show in Fig.\ \ref{simicska} the convergence of the 
phase shift $\delta_0(E)$ at a specific energy $E= 10$ MeV with and without 
the smoothing factors $\sigma_n^N$ in (\ref{1N}), and consequently 
in (\ref{v2b}). (Here and everywhere 
else the $\alpha$ parameter was chosen to be 5.2.) 
Clearly, the convergence is much poorer without 
the smoothing factors. We note that this also applies to the other 
quantities calculated for bound and resonance states.

Finally, we should call the attention upon the fact that this method is very
accurate. Reasonable accuracy is reached already at relatively small
basis, around $N=20$. The accuracy gained in larger bases 
is beyond most of the practical requirements. 
Test calculations have been performed on a linux PC (Intel PII, 266 MHz)
using double precision arithmetic. 
The calculation of a typical bound- or resonant-state 
energy requires the evaluation of the 
potential matrix by Gauss-Laguerre quadrature and finding  the zeros
of the determinant (\ref{determinans}), which incorporates 
the evaluation of the
Coulomb Green's matrix and the  the calculation of a determinant by 
performing an LU decomposition in each steps. 
The determination of the energy value in the first
column and last row of Table \ref{tab1}, which meant $6$ steps in the 
zero search and handling of $40\times 40$ matrices,
took $0.06$ sec. The corresponding resonance energy value in 
Table \ref{tab2} required $12$ steps in the zero search on the complex 
energy plane and  $0.8$ sec. 
The evaluation of the three phase shift values in the last row
of Table \ref{tab3} took $0.19$ sec., $0.11$ sec.\ and $0.08$ sec.,
respectively. 
So, this method is not only extremely accurate but also very fast.

\section{Conclusions}
\label{conc}

In this work we have presented a continued fraction representation of
the Coulomb Green's operator on Coulomb--Sturmian basis. 
Numerical illustrations show that this representations is simple, readily
computable and numerically exact on the whole complex energy plane.
This result is
based on the observation that in this basis the Coulomb Hamiltonian
possesses a Jacobi matrix structure. These techniques can be transferred  
to other problems, where the Hamiltonian matrix also has a Jacobi 
form. Examples for this are the harmonic oscillator \cite{jmp}, the 
generalized Coulomb problem \cite{jmp98}, which contains both the $D$ 
dimensional harmonic oscillator and the Coulomb problems as special 
limits and the relativistic Coulomb problem \cite{jmp99}. 

Once the representation of the Coulomb Green's operator 
in the discrete CS basis is available, we can proceed 
to solve   the Lippmann--Schwinger integral
equation. In practice this means 
the approximation of the potential term on a finite subset of this basis. 
This is the only stage where approximations are made, otherwise 
this method is exact and analytic, and provides asymptotically correct 
solutions. Consequently, bound, resonance and 
scattering problems can be treated on an equal footing, while 
these phenomena are usually discussed in rather 
different ways in conventional quantum mechanical approaches. 
This unified treatment is also reflected by the fact that all the 
calculations are made using the same discrete basis, containing 
also the same basis and other parameters.  
Furthermore, this approach is applicable to a wide range of 
potential problems, including complex, momentum-dependent, non-local, etc. 
ones. These techniques were illustrated with calculations for a 
model nuclear potential describing the interaction of two 
$\alpha$ particles. 

Once a discrete Hilbert-space basis representation of the Green's operator 
of the free or Coulomb Hamiltonian is at our disposal, 
we can also construct the
matrix representation of the Green's operator of the total Hamiltonian
by solving a resolvent equation. This provides a complete description
of the two-body system. The Green's operators, similarly to the wave 
functions, contain all the information about the system, and any physical 
property can be derived from them. The importance of this result is not 
purely formal, but it provides us with an effective tool in 
practical calculations for realistic physical problems. 
Moreover, from the Green's matrices of simpler systems, the Green's 
matrices of more complex composite systems can be constructed.
These techniques have been used so far for solving Faddeev equations of 
some three-body Coulombic systems \cite{pzwp,pzsc} but they can certainly be 
adapted to other challenging problems as well.

\section*{Acknowledgments}
B.~K.\ and Z.~P.\ are indebted to W.~Plessas for the kind hospitality and
useful discussions.
This work was supported by the  OTKA grants No.\ F20689,  No.\ T026233 
and No.\ T029003, and partially by the Austrian-Hungarian
Scientific-Technical Cooperation within project A-14/1998.

\begin{table}
\caption{Convergence of the $l=0$ bound-state energy eigenvalues $E_{nl}$ 
in $V_{\alpha-\alpha}(r)$ in the uncharged ($Z=0$) case. $N$ denotes  
the number of basis states used in the expansion. }
\begin{center}
\begin{tabular}{rccc}
$N$ & $E_{00}$ (MeV) & $E_{10}$ (MeV) & $E_{20}$ (MeV) \\
\hline \\
 8 & $-$76.903\ 557\ 1529 & $-$29.005\ 234\ 9134 & $-$1.739\ 478\ 2626 \\  
10 & $-$76.903\ 609\ 9717 & $-$29.000\ 352\ 3141 & $-$1.637\ 269\ 0831 \\  
15 & $-$76.903\ 614\ 3090 & $-$29.000\ 469\ 8249 & $-$1.608\ 824\ 6403 \\  
18 & $-$76.903\ 614\ 3254 & $-$29.000\ 470\ 2338 & $-$1.608\ 742\ 5166 \\ 
20 & $-$76.903\ 614\ 3263 & $-$29.000\ 470\ 2566 & $-$1.608\ 741\ 0685 \\  
25 & $-$76.903\ 614\ 3265 & $-$29.000\ 470\ 2623 & $-$1.608\ 740\ 8256 \\  
28 & $-$76.903\ 614\ 3265 & $-$29.000\ 470\ 2625 & $-$1.608\ 740\ 8216 \\ 
30 & $-$76.903\ 614\ 3265 & $-$29.000\ 470\ 2625 & $-$1.608\ 740\ 8213 \\  
35 & $-$76.903\ 614\ 3265 & $-$29.000\ 470\ 2626 & $-$1.608\ 740\ 8214 \\
40 & $-$76.903\ 614\ 3265 & $-$29.000\ 470\ 2626 & $-$1.608\ 740\ 8214  \\
\\
\end{tabular}
\label{tab1}
\end{center}
\end{table}

\begin{table}
\caption{
Convergence of the energy eigenvalues $E_{{\rm res},l}$ for the $l=0$ 
and $l=2$ resonances in the $V_{\alpha-\alpha}(r)$ potential. 
$N$ denotes the number of basis states used in the expansion. }
\begin{center}
\begin{tabular}{rccc}
$N$ & $E_{{\rm res}, 0}$  (MeV) & $E_{{\rm res}, 2}$ (MeV) \\
\hline \\
 8 & $-$0.000\ 854\ 9596  $+$i\ 0.000\ 000\ 0000  & 2.807\ 21 $-$i\ 0.607\ 11  \\  
10 &  \phantom{$-$}0.063\ 364\ 2503  $-$i\ 0.000\ 000\ 0681  & 2.866\ 30  $-$i\ 0.628\ 56
\\  
15 &  \phantom{$-$}0.091\ 785\ 0787  $-$i\ 0.000\ 002\ 8092  & 2.889\ 68  $-$i\ 0.620\ 99 
\\  
18 &  \phantom{$-$}0.091\ 963\ 0277  $-$i\ 0.000\ 002\ 8572  & 2.889\ 34  $-$i\ 0.620\ 53 
\\ 
20 &  \phantom{$-$}0.091\ 969\ 7296  $-$i\ 0.000\ 002\ 8588  & 2.889\ 24  $-$i\ 0.620\ 56  
\\  
25 &  \phantom{$-$}0.091\ 971\ 8479  $-$i\ 0.000\ 002\ 8592  & 2.889\ 23  $-$i\ 0.620\ 62  
\\  
28 &  \phantom{$-$}0.091\ 971\ 9788  $-$i\ 0.000\ 002\ 8592  & 2.889\ 25  $-$i\ 0.620\ 62  
\\ 
30 &  \phantom{$-$}0.091\ 972\ 0064  $-$i\ 0.000\ 002\ 8592  & 2.889\ 25  $-$i\ 0.620\ 61 
\\  
35 &  \phantom{$-$}0.091\ 972\ 0258  $-$i\ 0.000\ 002\ 8592  & 2.889\ 24  $-$i\ 0.620\ 61  
\\
40 &  \phantom{$-$}0.091\ 972\ 0290  $-$i\ 0.000\ 002\ 8592  & 2.889\ 25  $-$i\ 0.620\ 61 
\\
\\
\end{tabular}
\label{tab2}
\end{center}
\end{table}

\begin{table}
\caption{
Convergence of the $\delta_0(E)$ phase shift (in radians) 
in the $V_{\alpha-\alpha}(r)$ potential at three different energies. 
$N$ denotes the number of basis states used in the expansion. 
}
\begin{center}
\begin{tabular}{rccc}
$N$ & $E=0.1$ MeV & $E=1$ MeV & $E=30$ MeV \\
\hline \\
 8 & 6.283\ 230  & 8.817\ 731 & 7.783\ 217 \\  
10 & 9.424\ 059  & 8.862\ 581 & 4.817\ 163 \\  
15 & 9.424\ 018  & 8.859\ 651 & 4.835\ 479 \\  
18 & 9.424\ 022  & 8.859\ 467 & 4.829\ 861 \\ 
20 & 9.424\ 023  & 8.859\ 441 & 4.828\ 882 \\  
25 & 9.424\ 024  & 8.859\ 419 & 4.828\ 563 \\  
28 & 9.424\ 024  & 8.859\ 414 & 4.828\ 555 \\ 
30 & 9.424\ 024  & 8.859\ 412 & 4.828\ 554 \\  
35 & 9.424\ 024  & 8.859\ 411 & 4.828\ 552 \\
40 & 9.424\ 024  & 8.859\ 411 & 4.828\ 552 \\
\\
\end{tabular}
\label{tab3}
\end{center}
\end{table}

\begin{thebibliography}{99}

\bibitem{revai} J.~R\'evai, 
             JINR Preprint E4-9429, Dubna, 1975.
\bibitem{hopse} 
F.~A.~Gareev, M.~Ch.~Gizzatkulov, J.~R\'evai,  
             Nucl.\ Phys.\ {\bf A 286} 512 (1977); 
E.~Truhlik,  Nucl.\ Phys.\ {\bf A 296} 134 (1978); 
F.~A.~Gareev, S.~N.~Ershov, J.~R\'evai, J.~Bang,
             B.~S.~Nillsson,  Phys.\ Scripta {\bf 19}, 509 (1979); 
B.~Gyarmati, A.~T.~Kruppa, and J.~R\'evai, 
             Nucl.\ Phys.\ A {\bf 326}, 119 (1979); 
B.~Gyarmati, A.~T.~Kruppa, 
             Nucl.\ Phys.\ A {\bf 378}, 407 (1982);
B.~Gyarmati, A.~T.~Kruppa, Z.~Papp, and G.~Wolf,  Nucl.\ Phys.\ A 
             {\bf 417}, 393 (1984); 
A.~T.~Kruppa and Z.~Papp, 
             Comp. Phys. Comm. {\bf 36}, 59 (1985); 
J.~R\'evai, M.~Sotona, and J.~\v{Z}ofka,  J.\ Phys.\ G: Nucl.\ Phys.\ 
             {\bf 11}, 745 (1985); 
K.~F.~P\'al,   J.\ Phys.\ A: Math.\ Gen.\ {\bf 18}, 1665 (1985). 

\bibitem{papp1}  Z.~Papp,  J.\ Phys.\ A {\bf 20}, 153 (1987).

\bibitem{papp2}  Z.~Papp,  Phys.\ Rev.\ C 
	     {\bf 38}, 2457 (1988).

\bibitem{papp3} Z. Papp,  Phys.\ Rev.\ A {\bf 46}, 4437 (1992).

\bibitem{cpc}  Z.~Papp,  Comp.\ Phys.\ Comm.\
	     {\bf 70}, 426  (1992); ibid. 
	     {\bf 70}, 435 (1992).

\bibitem{pzwp} Z.~Papp and  W. Plessas,  
             Phys.\ Rev.\ C, {\bf 54}, 50  (1996);  
             Z.\ Papp,  
             Few-Body Systems, {\bf 24} 263   (1998).

\bibitem{pzsc} Z. Papp,   
	     Phys.\ Rev.\ C, {\bf 55}, 1080  (1997).

\bibitem{jmp} B.~K\'onya, G.~L\'evai and Z.~Papp,  
             J.\ Math.\ Phys.\ {\bf 38}, 4832  (1997).

\bibitem{jones}  W.~B.~Jones and W.~J.~Thron, {\it  Continued Fractions:
                 Analytic Theory and Applications} (Addison-Wesley,
                 Reading, 1980).

\bibitem{lorentzen}  L.~Lorentzen and H.~Waadeland, {\it  Continued
               Fractions with Applications} (Noth-Holland, Amsterdam,  
               1992).

\bibitem{rotenberg} M.~Rotenberg,  
               Ann.\ Phys.\ (N.Y.) {\bf 19}, 262  (1962); \\
               M.~Rotenberg,  
	       Adv.\ At.\ Mol.\ Phys.\ {\bf 6}, 233  (1970).

\bibitem{as} M.~Abramowitz and I.~Stegun, {\it Handbook of Mathematical
                 Functions} (Dover, New York, 1970).
                 
\bibitem{newton}  R.~G.~Newton, {\it Scattering Theory of Waves and
                  Particles} (Springer, New York, 1982).

\bibitem{lanczos} C. Lanczos, {\it Linear Differential Operators} 
                (D.~van Nostrand, London, 1961). 

\bibitem{borbely} I.~Borb\'ely, private communication in Ref.\cite{revai},
1975. The $\sigma$
factors were subsequently used in all calculations of Refs.\ 
\cite{revai,hopse,papp1,papp2,papp3,cpc,pzwp,pzsc}.

                 
\bibitem{schmid} E.~W.~Schmid, G. Spitz and W. L\"osch, 
                {\it Theoretical Physics on the Personal Computer} 
                (Springer, Berlin, 1988), Chapter 13.
		
\bibitem{ajzenberg} F.~Ajzenberg-Selove, Nucl.\ Phys.\ A {\bf 490}, 1
 (1988).

\bibitem{jmp98} G.~L\'evai, B.~K\'onya and Z.~Papp, 
 J.\ Math.\ Phys.\ {\bf 39}, 5811 (1998). 

\bibitem{jmp99} B.~K\'onya and Z.~Papp,  
                J.\ Math.\ Phys.\ {\bf 40}, 2307 (1999).

                           
\end{thebibliography}
\end{document}